\documentclass[twocolumn,astrosymb,preprint2,twocolappendix]{aastex631}
\usepackage{amsmath}
\usepackage{bm}
\usepackage{xcolor}
\newcommand{\ubb}{\ensuremath{\boldsymbol{u}}}
\newcommand{\xb}{\ensuremath{\boldsymbol{x}}}
\newcommand{\yb}{\ensuremath{\boldsymbol{y}}}
\newcommand{\zb}{\ensuremath{\boldsymbol{z}}}
\newcommand{\nb}{\ensuremath{\boldsymbol{n}}}
\newcommand{\Fb}{\ensuremath{\boldsymbol{\mathsf{F}}}}
\newcommand{\Gb}{\ensuremath{\boldsymbol{\mathsf{G}}}}
\newcommand{\Zb}{\ensuremath{\boldsymbol{\mathsf{Z}}}}
\newcommand{\Hb}{\ensuremath{\boldsymbol{\mathsf{H}}}}
\newcommand{\Psib}{\ensuremath{\boldsymbol{\Psi}}}
\newcommand{\Phib}{\ensuremath{\boldsymbol{\Phi}}}
\newcommand{\eC}{\mathbb{C}}
\newcommand{\eN}{\mathbb{N}}
\newcommand{\eR}{\mathbb{R}}
\usepackage{xcolor}
\shorttitle{AI for deep wide-field imaging}
\shortauthors{Dabbech et al.}
\graphicspath{{./}{}}

\begin{document}
\title{First AI for deep super-resolution wide-field imaging in radio astronomy: unveiling structure in ESO~137--006}

\correspondingauthor{A. Dabbech}
\email{a.dabbech@hw.ac.uk}
\author[0000-0002-7903-3619]{A.~Dabbech}
\affiliation{Institute of Sensors, Signals and Systems, Heriot-Watt University,
Edinburgh EH14 4AS, UK}
\author{M. Terris}
\affiliation{Institute of Sensors, Signals and Systems, Heriot-Watt University,
Edinburgh EH14 4AS, UK}
\author[0000-0003-0073-682X]{A. Jackson}
\affiliation{EPCC, University of Edinburgh,  Edinburgh EH8 9BT, UK}
\author[0000-0003-0231-3249]{M. Ramatsoku}
\affiliation{Department of Physics and Electronics, Rhodes University, PO Box 94, Makhanda (Grahamstown), 6140, South Africa}
\author[0000-0003-1680-7936]{O.~M. Smirnov}
\affiliation{Department of Physics and Electronics, Rhodes University, PO Box 94, Makhanda (Grahamstown), 6140, South Africa}
\affiliation{South African Radio Astronomy Observatory, 2 Fir Street, Black River Park, Observatory, Cape Town 7925, South Africa}
\author[0000-0002-1658-0121]{Y. Wiaux}
\affiliation{Institute of Sensors, Signals and Systems, Heriot-Watt University,
Edinburgh EH14 4AS, UK}



\begin{abstract}

We introduce the first AI-based framework for deep, super-resolution, wide-field radio-interferometric imaging, and demonstrate it on observations of the ESO~137-006 radio galaxy. The algorithmic framework to solve the inverse problem for image reconstruction builds on a recent ``plug-and-play'' scheme whereby a denoising operator is injected as an image regulariser in an optimisation algorithm, which alternates until convergence between denoising steps and gradient-descent data-fidelity steps. We investigate handcrafted and learned variants of high-resolution high-dynamic range denoisers. We propose a parallel algorithm implementation relying on automated decompositions of the image into facets and the measurement operator into sparse low-dimensional blocks, {enabling scalability to large data and image dimensions. We validate our framework for image formation} at a wide field of view containing ESO~137-006, from 19 gigabytes of MeerKAT data at 1053 and 1399 MHz. The recovered maps exhibit significantly more resolution and dynamic range than CLEAN, revealing collimated synchrotron threads close to the galactic core.

\end{abstract}



\section{Introduction} \label{sec:intro}

Image formation in aperture synthesis by radio interferometry (RI) has never been more challenging. On the one hand, extreme data sampling rates, produced by modern radio arrays, raise the urgent need for scalable algorithms. On the other hand, the ill-posedness of the underlying inverse problem calls for tailored regularisation models to be injected in the image formation process in order to deliver the expected precision and robustness of the reconstruction. Over the last decade, regularisation approaches leveraging advanced sparsity-based image models embedded in optimisation algorithms were proposed \citep[\emph{e.g.}][]{Li2011,Carrillo2012,Dabbech2015,Garsden2015}. In particular, the SARA family of algorithms \citep{onose2017accelerated,repetti17,dabbech18,Abdulaziz19,dabbech21, thouvenin2022parallel1,thouvenin2022parallel2} have recently delivered a significant increase of resolution and dynamic range (or depth) over CLEAN-based algorithms \citep[\emph{e.g.}][]{hogbom74,Wakker88,cornwell08b} on the well known radio galaxy Cygnus~A. Owing to the complexity of the regularisation models underpinning the imaging accuracy, optimisation approaches are  significantly more computationally expensive than CLEAN. Their scalability to gigabyte-scale image size and data volumes has been enabled by resorting to advanced algorithmic structures enabling a significant degree of parallel processing \citep[\emph{e.g.}][]{pesquet14,Chouzenoux15}. Nonetheless, scalability to much larger image and data dimensions is required for upcoming instruments, with the Square {Kilometre} Array (SKA) \citep{Scaife20} intended to deliver petabyte-scale images from exabyte-scale data volumes.

Assuming monochromatic, non-polarised radio emission, and a narrow field-of-view (FoV), the measured complex visibilities are noisy Fourier components of the sky surface brightness, where the sampled $(u,v)$ points are the projections of each antenna pair baseline on the plane perpendicular to the line of sight. Under these assumptions, the visibility vector $\yb \in \eC^M$ can be modelled as
\begin{equation}
\label{eq:invpb}
 \yb = {\Phib}\overline{\xb}+ \overline{\nb}, \textrm{~with~} {\Phib} = \Gb\Fb{\Zb},
\end{equation}
where $\overline{\xb}\in \eR^N$ is the unknown radio map, whose pixel resolution is often set between 1.5 to 2.5 times below the angular resolution of the observations, to reduce the limitations of pixel-based image restoration. $\overline{\nb}\in \eC^M$ is a realisation of a complex random Gaussian noise of mean 0 and standard deviation $\tau>0$. $\Phib\in \eC^{M \times N}$ denotes the measurement operator, which encodes the incomplete Fourier sampling. More precisely, $\Gb\in \eC^M \times \eC^{N^\prime}$ denotes a sparse de-gridding matrix whose rows are non-uniform Fourier transform interpolation kernels, $\Fb\in \eC^{N^\prime} \times \eC^{N^\prime}$ stands for the Discrete Fourier transform, and $\Zb\in \eC^{N^\prime}\times \eR^{N} $ is a zero-padding operator, allowing for a fine grid in the spatial Fourier domain, which also involves a correction for approximations in the convolution kernels of $\Gb$ \citep{Fessler2003}. Given the remarkable sensitivity of the modern arrays, the RI measurement equation is further complicated by the so-called direction-dependent effects (DDEs). Some of these are unknown and of either atmospheric or instrumental origin and should be calibrated \citep{Smirnov2011}. 
In contrast, the DDEs originating from the $w$ component of the antenna pair baselines on the line of sight are known, and induce the so-called $w$-effect \citep{Cornwell1992}. 
DDEs can be encapsulated as additional baseline-specific convolution kernels on each row of $\Gb$ \citep{dabbech17,repetti17}. In this work, we propose a new parallelised and automated framework for wide-field high-resolution high-dynamic range monochromatic intensity imaging, which we use to revisit observations of the radio galaxy ESO~137-006, the loudest radio galaxy in the Norma cluster.

{The remainder of this letter is structured as follows. In section \ref{sec:mainmethods}, we provide a summary of the proposed framework, from the underpinning algorithmic structure and the two specific incarnations respectively propelled by sparsity-based and AI-based regularisation, to parallelisation and automation functionalities critical to scalability. 
A description of the utilised RI data of ESO~137 from the MeerKAT telescope is given in section \ref{sec:maindata}. Imaging settings as well as a description of the utilised computational resources are provided in section \ref{sec:imagingset}. Imaging results are presented in comparison with a CLEAN-based benchmark method in section \ref{sec:mainres}, followed by a discussion on the unveiled structure in ESO~137-006. Finally, conclusions are drawn in section \ref{sec:mainconcl}.}

\section{Methods}
\label{sec:mainmethods}
At the algorithmic level, the {proposed} framework is underpinned by the versatile Forward-Backward (FB) convex optimisation iterative structure \citep{bauschke2017convex}. At each iteration, FB simply alternates until convergence between a (forward) gradient-descent step promoting fidelity to data and a (backward) step enforcing a prior image model, critical to the regularisation of the inverse problem and the resulting imaging precision {(see appendix \ref{sec:fb})}. We investigate two incarnations of a recent plug-and-play (PnP) scheme \citep{venkatakrishnan2013plug, romano2017little}, whereby  dedicated denoising operators can be plugged into FB as an image regulariser.

The unconstrained SARA (uSARA) algorithm is a pure optimisation variant leveraging a so-called ``proximal'' denoiser, handcrafted to enforce an advanced sparsity-based image regularisation \citep{Carrillo2012,repetti21,terris22}. The sophistication of the underlying prior image model is precisely introduced to deliver the best possible resolution and dynamic range from the data. The resulting denoiser itself is implemented as an iterative algorithm, leading to an overall sub-iterative FB structure {(see appendix \ref{ssec:sara})}. 

The AIRI (standing for ``AI for Regularisation in radio-interferometric Imaging'') algorithm \citep{terris22} is an AI-based variant leveraging a learned denoiser in the form of a deep neural network (DNN) trained on a rich database to clean Gaussian random noise from high dynamic range images, with a noise level commensurate with the target sensitivity of observation {(see appendix \ref{ssec:airi})}.  By design, AIRI inherits the robustness and interpretability of optimisation algorithms and the learning power and speed of DNNs. 

Importantly, the degree of refinement with which the uSARA and AIRI image models are enforced is adjusted to the measurement noise $\tau$, more precisely to the corresponding estimate of the noise level in the image domain, $\tau/\sqrt{2L}$, which results from a normalisation by the norm $L$ of the measurement operator. In other words, uSARA and AIRI automatically adapt to the input signal-to-noise ratio, or equivalently, the target dynamic range of reconstruction. Last but not least, we emphasise that, by construction, PnP denoisers are completely blind to the measurement conditions underpinning the data to be imaged. As a consequence, the learned variants can be trained once and for all at an appropriate dynamic range, significantly alleviating the associated computation cost. They do not suffer from generalisability challenges with respect to measurement conditions either. This stands in stark contrast with the more traditional end-to-end approaches, where a DNN would be trained to reconstruct an image directly from data \citep{Connor22,terris22}.

At the high-resolution and high-dynamic range regime of interest, 
parallelisation and automation functionalities are critical to the scalability of the algorithmic framework. In this context, the image denoisers of uSARA and AIRI are decomposed on small image facets with no loss of precision thanks to their convolutional nature and the compactness of the associated kernels {(see appendix \ref{app:denoiser})}. Relying on a hybrid approach to efficiently correct for the wide-field $w$-effect in both image and data spaces \citep{Cornwell2005,Wiaux09,Offringa2014,dabbech17}, the measurement operator is decomposed into sparse and low-dimensional building blocks {(see appendix \ref{sec:mo})}. These decompositions are fully automated, enabling a parallel image facet and data block processing, seamlessly adapting to the architecture of the high performance computing (HPC) system where the reconstruction is run.

\section{Data description}
\label{sec:maindata}
Both uSARA and AIRI are used to revisit MeerKAT L-band observations of a wide FoV containing the radio galaxy ESO~137-006. MeerKAT \citep{Jonas2016}, located in the Karoo desert of South Africa, is a precursor to the SKA. Its 64 antennas with cryogenic receivers are arranged in a close-packed core and baselines of up to 8 kilometres, resulting in superb sensitivity and imaging quality. The array is particularly suited to study faint extended emission and objects with complex morphology, of which ESO~137-006 represents a ``flagship'' case. Previous analysis of these observations by \citet{ramatsoku20} revealed multiple collimated synchrotron threads (CSTs) connecting the lobes of the radio galaxy, whose origin is yet to be unravelled.

Technical details of the observations and the initial calibration (\emph{i.e.}, reference calibration or 1GC) is reported by \citet{ramatsoku20}. 1GC was performed using the CARACAL pipeline \citep{makhathini2018,MeerKAThi}. The 1GC-calibrated data are averaged down from 4096 to 1024 channels of 0.84 MHz each, spanning the frequency range 856--1712 MHz. 
We utilise about 7 hours of on-target time and select two sub-bands relatively free from {radio frequency interference}, referred to as the ``low'' band (961-1145 MHz, centred at 1053~MHz) and the ``high'' band (1295-1503 MHz, centred at 1399~MHz), to form two continuum images. The respective data sizes after flagging are 8.2 gigabytes ($\sim$532 million data points, double precision) and 10.76 gigabytes ($\sim$673 million data points). The data were then self-calibrated for phase using a combination of the WSClean imager \citep{offringa17} and the CubiCal calibration suite \citep{Kenyon18}. {Attempts to  calibrate for the amplitude and the DDEs \citep{repetti17, dabbech21} did not bring a substantial improvement. Therefore, no further data pre-processing was performed}. The resulting WSClean images, obtained with the multiscale variant of CLEAN \citep{cornwell08b}, are presented for comparison purposes. uSARA and AIRI are then used for image reconstruction on these self-calibrated data. 

\section{Imaging settings $\&$ computational resources}
\label{sec:imagingset}
The images formed are of size $4096\times 4096$ pixels, spanning the FoV $1.91 \times 1.91$ square degrees with a cell size of $1.68$ arcseconds, for super-resolution factors beyond the angular resolution of observation of about 2 and 1.6 at low and high bands respectively. Data were weighted using the Briggs weighting scheme (robust parameter 0) to mitigate at best the complicated lobes of the dirty beam, \emph{i.e.}~the point spread function arising from the Fourier sampling pattern. Specifically to AIRI, a single denoiser with appropriate dynamic range was trained and used as AIRI regulariser at both bands. Imaging parameter selection for both uSARA and AIRI is automated (see appendix \ref{app:params}). {Finally, WSClean parameters are set similarly to \citet{ramatsoku20}.}

With regards to computing resources, the MATLAB implementation of uSARA and AIRI and the C++ WSClean imager were run {on Cirrus\footnote{\href{http://www.cirrus.ac.uk}{http://www.cirrus.ac.uk}}}, a UK Tier2 HPC
system. uSARA and CLEAN are deployed on CPUs, while AIRI is deployed mainly on CPUs, with AIRI's denoiser utilising a GPU. 
More precisely, for uSARA and AIRI, the {computation} of the measurement operator (see appendix~\ref{sec:mo} for details),
decomposed into sparse low-dimensional building blocks, utilised 240 and 280 CPUs at low and high bands respectively. For the imaging process itself, forward steps utilised 99 and 180 CPUs at low and high bands respectively.  
uSARA's denoiser, distributed over 64 image facets, utilised 64 of the CPUs already allocated for the forward steps. AIRI's denoiser relied on a decomposition of the image into 4 facets, lowering the memory requirements per facet and enabling each facet to be processed on a single GPU. Given the relatively negligible GPU computation cost, a single GPU was used, with facets denoised sequentially rather than in parallel. Finally, WSClean used 72 CPUs, associated with the considered number of $w$-stacks. %
\section{Results and discussion}
\label{sec:mainres}
\begin{figure*}[htb!]
\gridline{\fig{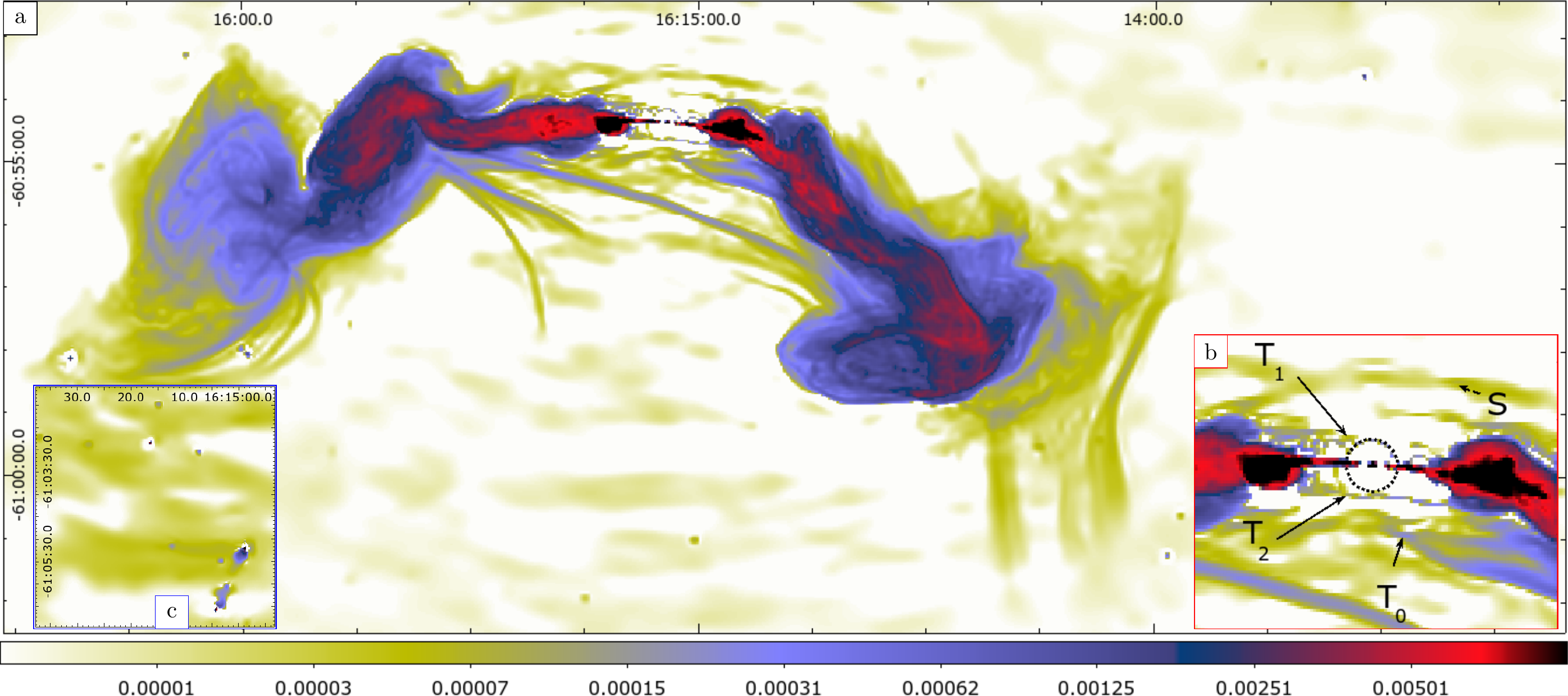}{.8\textwidth}{}}\vspace{-0.8cm}\label{fig:usara_im_hi}
\gridline{\fig{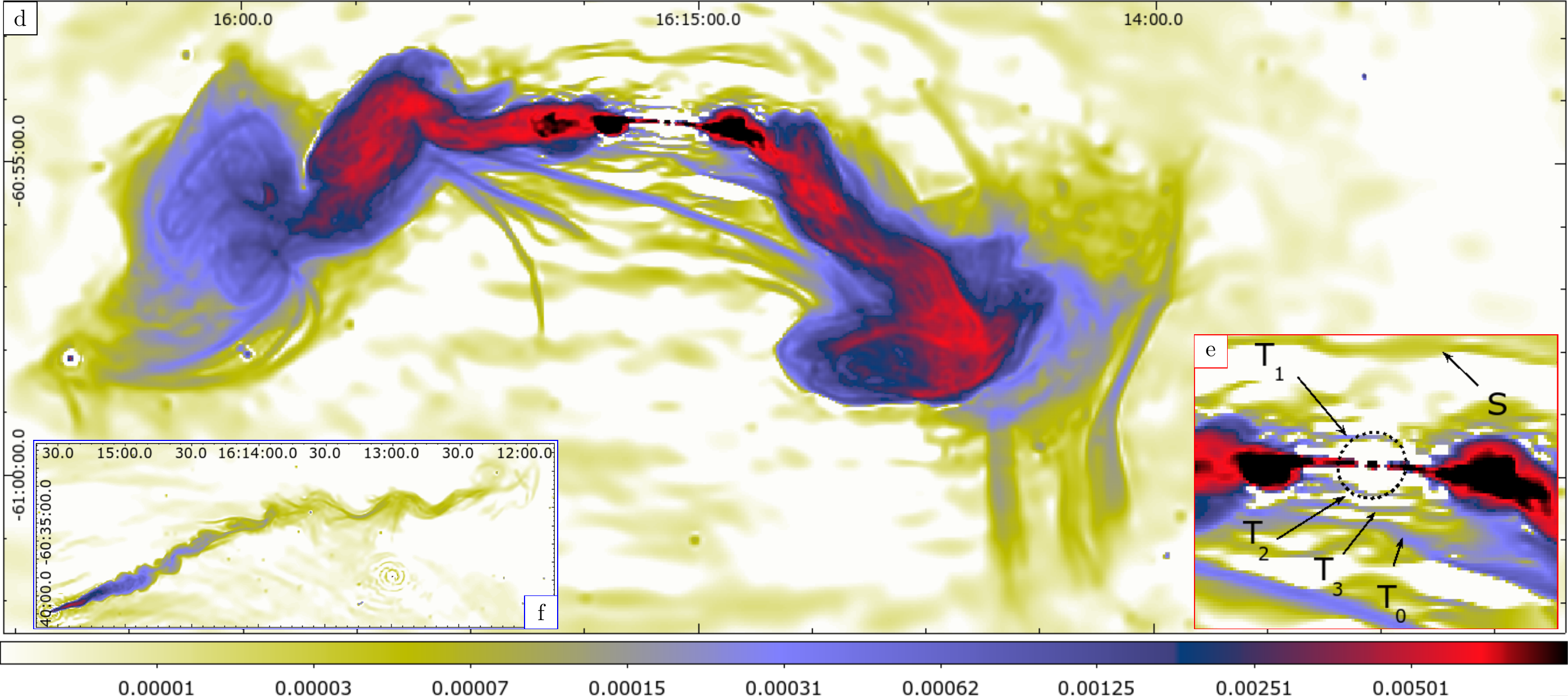}{.8\textwidth}{}}\vspace{-0.8cm}\label{fig:usara_im_low}
\gridline{\fig{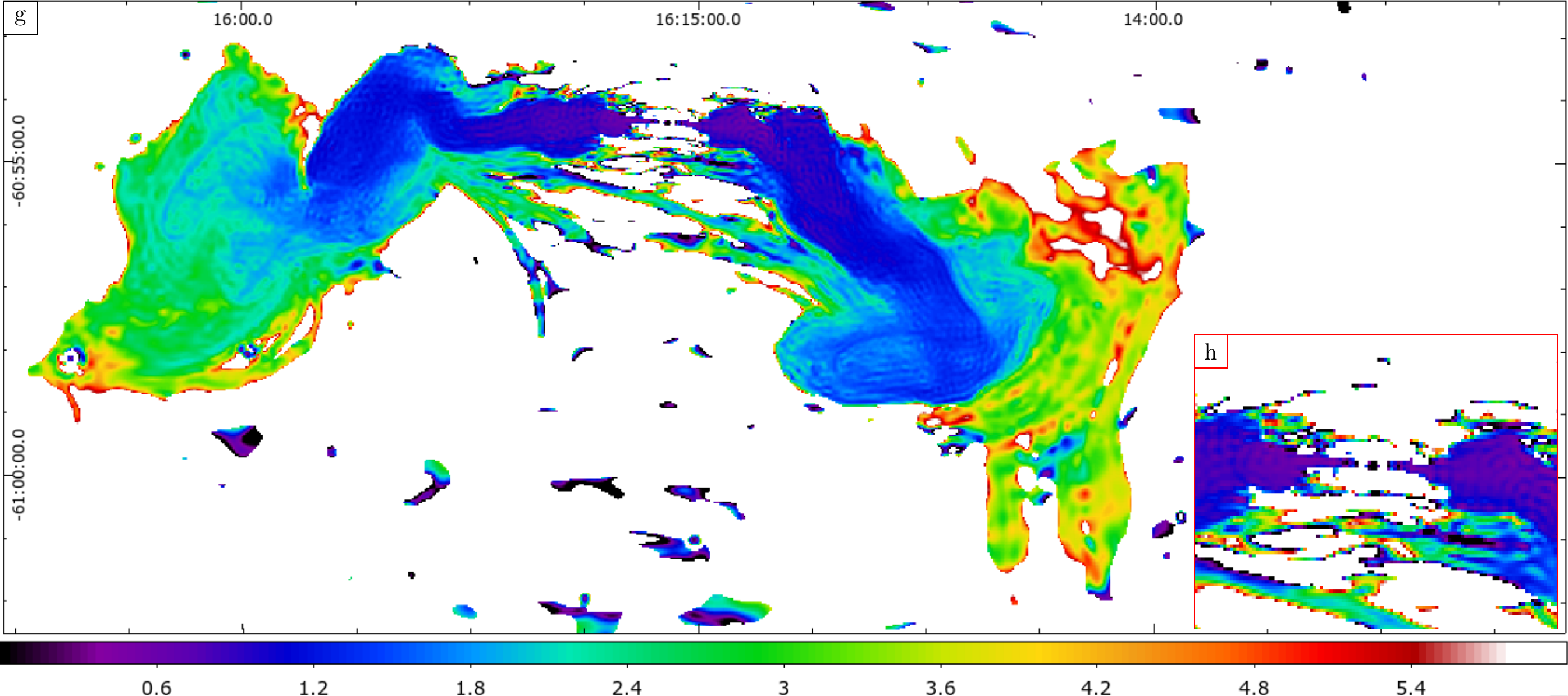}{0.8\textwidth}{}}\vspace{-0.8cm}\label{fig:usara_spi}
\caption{\footnotesize ESO~137--006: uSARA reconstructions (flip pages to visualise differences at a glance with AIRI in Figure \ref{fig:airi} and CLEAN in Figure \ref{fig:wsclean}). First and second rows: recovered model images (Jy/pixel, displayed in log$_{10}$ scale) at high and low bands (panels (a) and (d)), respectively, overlaid with zooms on the core of ESO~137--006 (panels (b) and (e)), a region with compact sources from the imaged FoV (panel (c)), and a zoom on ESO~137--007, a radio galaxy North of ESO~137--006 (panel (f)). Third row: spectral index map of ESO~137--006 (displayed in linear scale, panel (g)), overlaid with a zoom on its core (panel (h), same region as in panels (b) and (e)). Focusing on the central region (panels (b) and (e)), the first sidelobe of the dirty beam is highlighted with a dashed circle. One can see three filaments emerging: $\textrm{T}_1$ and $\textrm{T}_2$, located North and South of the inner core, seen at both bands, and $\textrm{T}_3$, located further South, recovered only at low band. A fourth filament $\textrm{T}_0$,
detected previously, is also recovered. The filamentary structure $\textrm{S}$ is an example of what mostly likely is a calibration residual artefact, as it moves with the geometry of the dirty beam. The spectral index values of the newly formed filaments (panel (h)) are inconclusive.
\label{fig:sara}}
\end{figure*}

\begin{figure*}[htb!]
\gridline{\fig{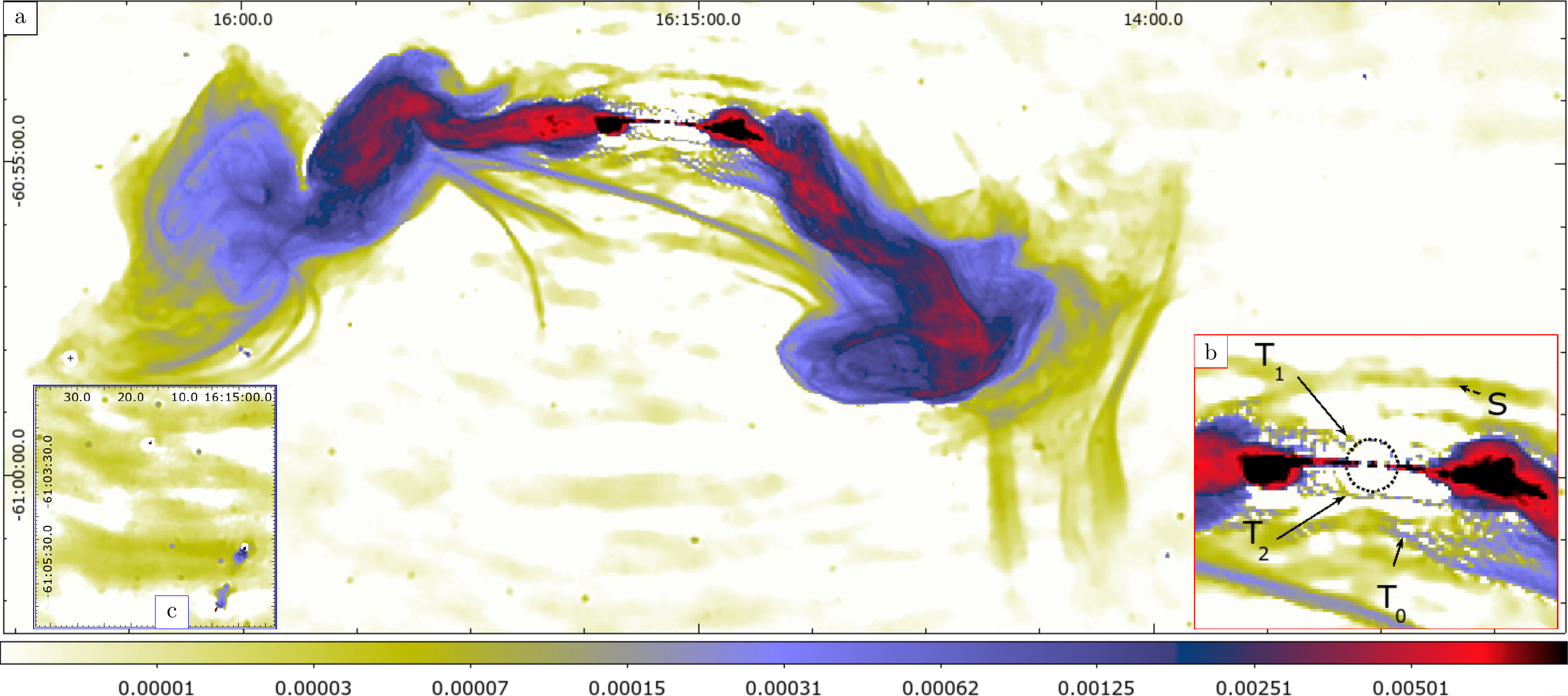}{.8\textwidth}{}}\vspace{-0.8cm}\label{fig:airi_im_hi}
\gridline{\fig{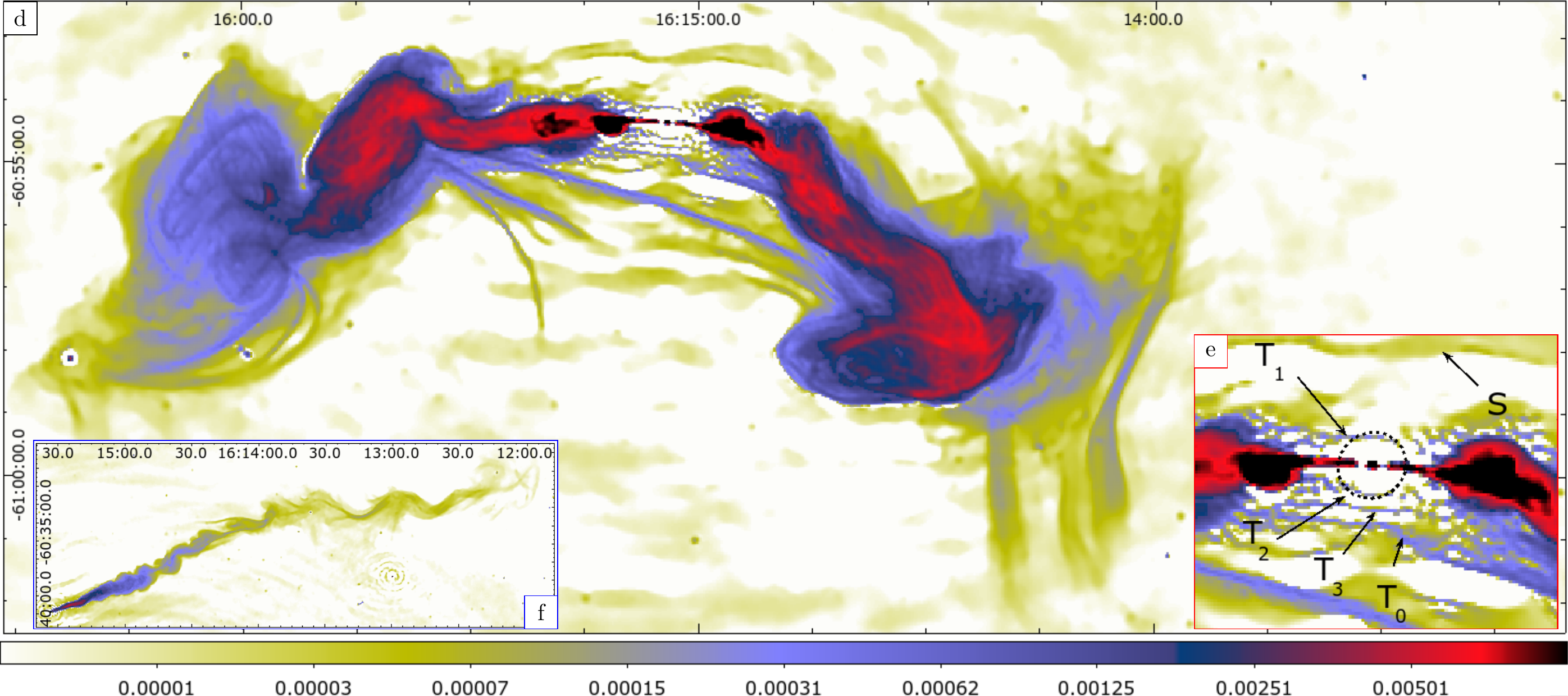}{.8\textwidth}{}}\vspace{-0.8cm}\label{fig:airi_im_low}
\gridline{\fig{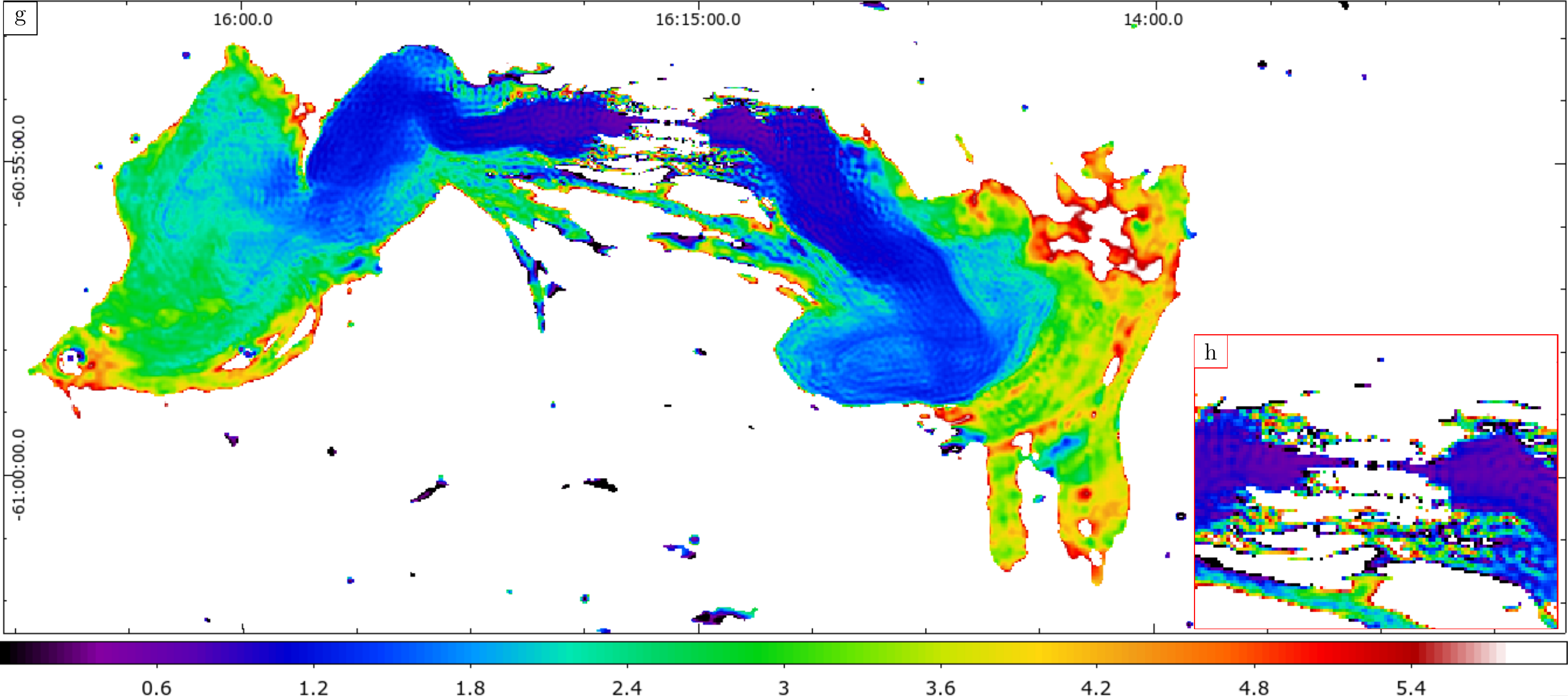}{0.8\textwidth}{}}\vspace{-0.8cm}\label{fig:airi_spi}
\caption{\footnotesize ESO~137--006: AIRI reconstructions (flip pages to visualise differences at a glance with uSARA in Figure \ref{fig:sara} and CLEAN in Figure \ref{fig:wsclean}). First and second rows: recovered model images (Jy/pixel, displayed in log$_{10}$ scale) at high and low bands (panels (a) and (d)), respectively, overlaid with zooms on the core of ESO~137--006 (panels (b) and (e)), a region with compact sources from the imaged FoV (panel (c)), and a zoom on ESO~137--007, a radio galaxy North of ESO~137--006 (panel (f)). Third row: spectral index map of ESO~137--006 (displayed in linear scale, panel (g)), overlaid with a zoom on its core (panel (h), same region as in panels (b) and (e)). Focusing on the central region (panels (b) and (e)), the first sidelobe of the dirty beam is highlighted with a dashed circle. Similarly to uSARA reconstructions, one can see three filaments emerging: $\textrm{T}_1$ and $\textrm{T}_2$, located North and South of the inner core, seen at both bands, and $\textrm{T}_3$, located further South, recovered only at low band. A fourth filament $\textrm{T}_0$, detected previously, is also recovered. The filamentary structure $\textrm{S}$ is an example of what mostly likely is a calibration residual artefact, as it moves with the geometry of the dirty beam. The spectral index values of the newly formed filaments (panel (h)) are inconclusive.\label{fig:airi}}
\end{figure*}

\begin{figure*}[htb!]
\gridline{\fig{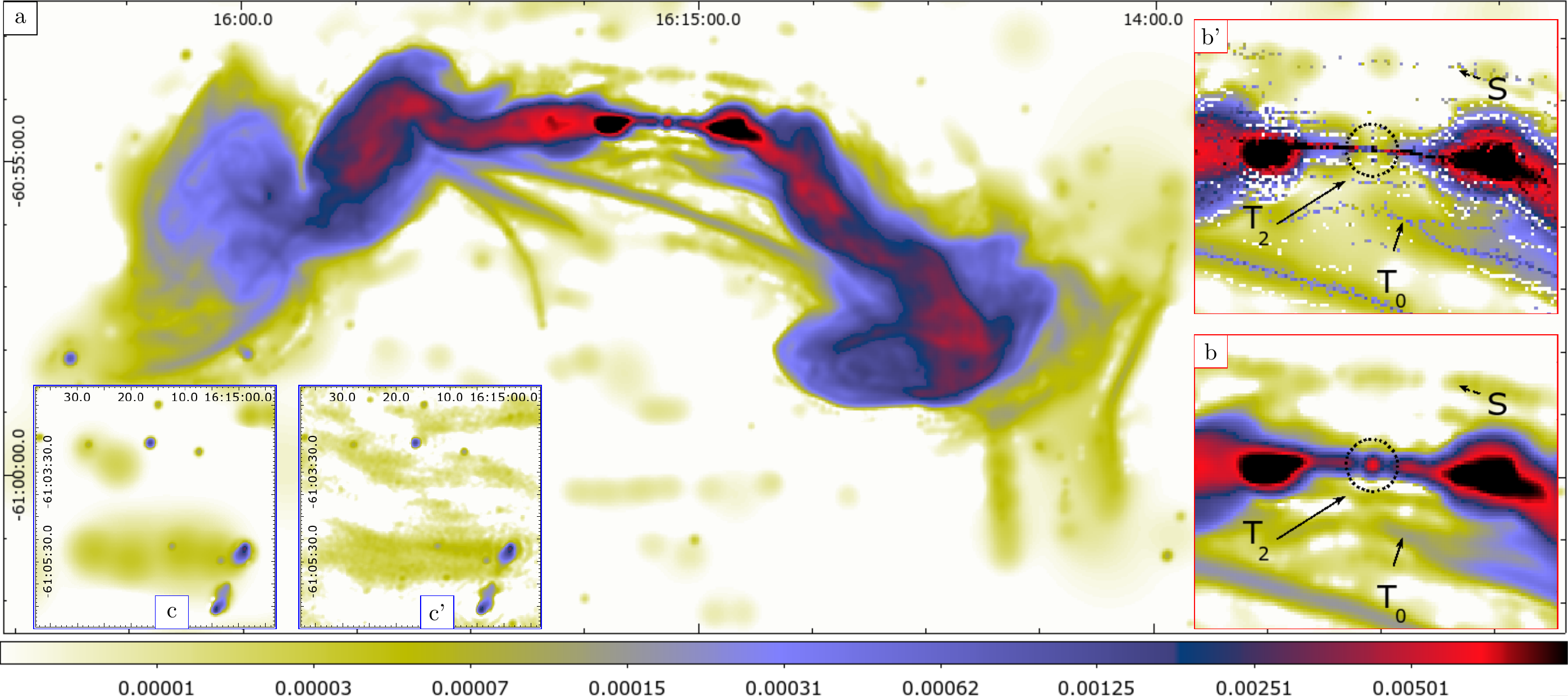}{.8\textwidth}{}}\vspace{-0.8cm}\label{fig:ws_im_hi}
\gridline{\fig{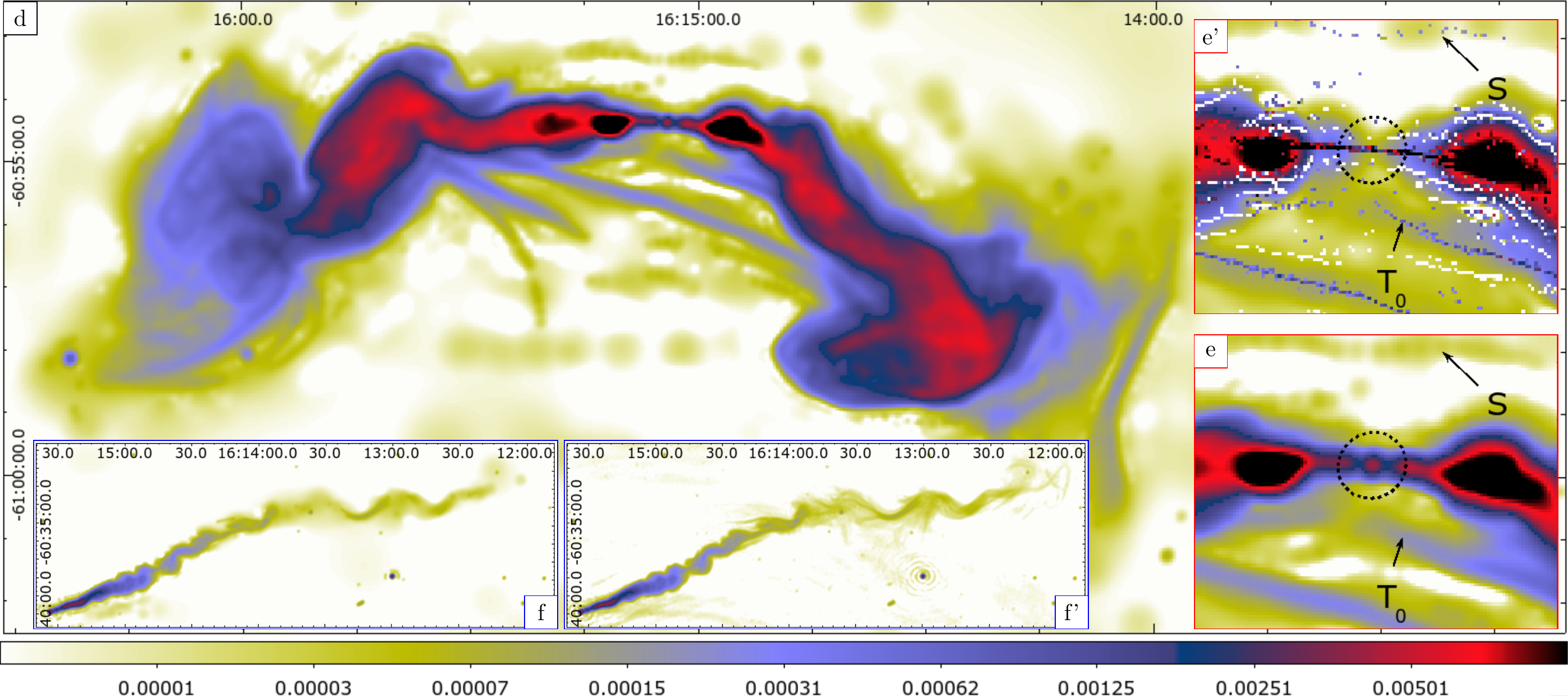}{.8\textwidth}{}}\vspace{-0.8cm}\label{fig:ws_im_low}
\gridline{\fig{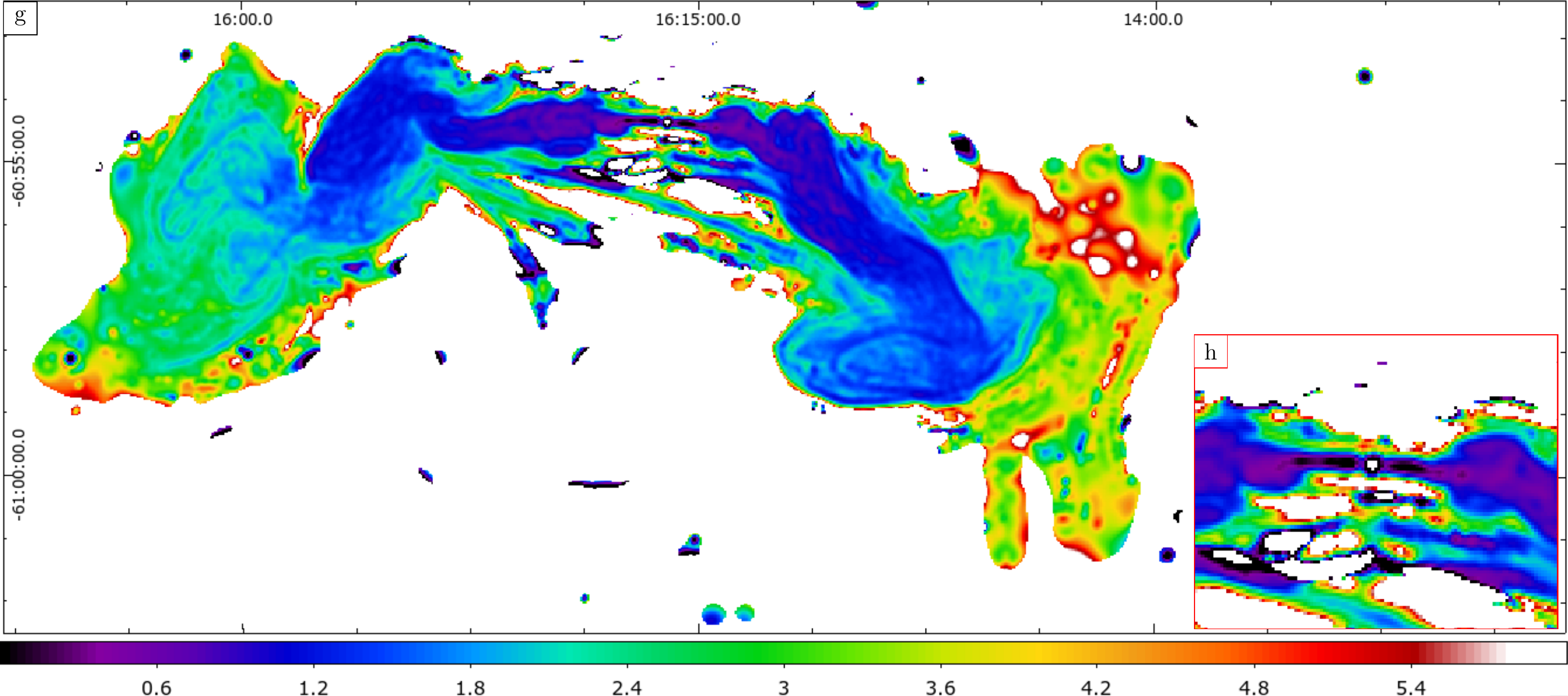}{0.8\textwidth}{}}\vspace{-0.8cm}\label{fig:ws_spi}
\caption{\footnotesize {ESO~137--006: CLEAN reconstructions (flip pages to visualise differences at a glance with uSARA in Figure \ref{fig:sara} and AIRI in Figure \ref{fig:airi}). First and second rows: recovered convolved model images (Jy/pixel, displayed in log$_{10}$ scale) at high and low bands (panels (a) and (d)), respectively, overlaid with zooms on the core (panels (b) and (e)), a region with compact sources from the imaged FoV (panels (c) and (c')), and a zoom on ESO~137--007, a radio galaxy North of ESO~137--006 (panels (f) and (f')). Third row: estimated spectral index map of ESO~137--006 (displayed in linear scale, panel (g)), overlaid with a zoom on its core (panel (h), same region as in panels (b){, (b')} and (e)). Firstly, regarding the central region, the first sidelobe of the dirty beam is highlighted with a dashed circle. Unlike uSARA and AIRI reconstructions, only one new filament $\textrm{T}_2$ has clearly emerged South of the core at high band. The previously detected filament $\textrm{T}_0$ is recovered at both bands. The filamentary structure $\textrm{S}$ is an example of what mostly likely is a calibration residual artefact, as it moves with the geometry of the dirty beam. {Inspection of the zooms on the core from CLEAN model images (panels (c') and (e')) confirms these findings.} Secondly, looking at the zooms on the compact sources, one can see that some faint sources are not recovered in the CLEAN model image (panel (c)). Instead, they are left in the residual, and so only visible in the CLEAN restored image (panel (c')). Finally, a close look at ESO~137--007 in the convolved model image (panel (f)), indicates that much of the filamentary structure is not captured, and is only visible on the CLEAN restored image (panel (f')).}\label{fig:wsclean}}
 \end{figure*}

Reconstruction results are provided in Figures~\ref{fig:sara}-\ref{fig:wsclean}, focusing on the ESO~137-006 region of the imaged FoV, and displayed in $\textrm{log}_{10}$ scale to enable the joint visualisation of high intensity and faint emission. Specifically to CLEAN reconstructions, we display the outcome of the convolution of the associated model image with the so-called restoring beam, for a more physical representation of the radio sky. By construction, uSARA and AIRI images are in units of $\textrm{Jy}/\textrm{pixel}$. In order to compare intensities in the same units, CLEAN images are normalised by the flux of the restoring beam. 
Zooms on selected regions of the imaged FoV are provided on each figure. Firstly, a zoom on the central region of ESO~137-006, including the active galactic nucleus (AGN) at its core, is provided in panels (b) and (e). Secondly, a zoom on some background compact sources at high band are shown in panel (c). Thirdly, a zoom on the neighbouring radio galaxy ESO~137--007 North of ESO~137--006, at low band, is displayed in panel (f). Images of the full FoV are provided as supplementary material \citep{eso137}. 

Where the residual images contain additional information, CLEAN's restored image, consisting of the sum of the convolved model and the residual image, is considered in our analysis. {Both zooms on the selected background compact sources at high band and the neighbouring radio galaxy ESO~137--007 at low band from CLEAN restored images are included in the respective panels (c') and (f') of Figure~\ref{fig:wsclean}}. No such consideration is necessary for uSARA and AIRI, where the algorithm solution itself, without further processing, is considered to be the final image reconstruction. This advantage was already highlighted for the previous algorithms of the SARA family, which rely on more advanced and physical regularisation models than CLEAN \citep{dabbech18,Abdulaziz19,dabbech21,thouvenin2022parallel2}. {Finally, the model image of CLEAN is considered to support our analysis, particularly through zooms of the central region of ESO~137-006 at high and low bands shown in the respective panels (b') and (e') of Figure~\ref{fig:wsclean}.}

Generally speaking, on both bands, one can observe the high level of detail achieved by uSARA and AIRI in comparison with CLEAN, particularly noticeable within the lobes of the radio galaxy. As opposed to the maps produced by CLEAN, whose resolution is, by design, restricted due to the convolution with the restoring beam,  uSARA and AIRI maps show a wealth of filamentary detail within the radio lobes of ESO~137--006. These improvements also come with some pixelation effects, more noticeable in AIRI's reconstructions, in super-resolved structures around the galactic core/jets.

The ability of both uSARA and AIRI to capture complex structure is further showcased when looking at the zoom on ESO~137--007 at low band (a similar observation is made at high band), in contrast with CLEAN (panel (f) of each figure). In fact, the filamentary detail of its jet is not recovered in the CLEAN convolved model image. Instead, it is left in the residual image and therefore only seen in the CLEAN restored image (panel (f') of Figure~\ref{fig:wsclean}). uSARA and AIRI also deliver further reconstruction depth, recovering lower signal intensities than CLEAN. {Additional examination of the bright super-resolved point-like sources South of the jet showcases the improvement in resolution brought by both AIRI and uSARA.} 
Last but not least, one can notice that AIRI further improves the reconstruction dynamic range over uSARA, as is apparent from the recovered faint compact sources at high band in panel (c) of each figure (a similar observation is made at low band). These sources are also visible in the CLEAN restored image (panel (c')), even though not directly recovered in the CLEAN model image (panel (c)). This supports the fact that they are real and not hallucination artefacts of the DNN. {AIRI also seems to be less sensitive to calibration errors which take the form of extended ringing structure around ESO~137--006 (panels (a) and (d) of Figures~\ref{fig:sara}-\ref{fig:airi})}.

Since the low and high band images are produced completely independently, flux measurements of unresolved sources provide an important cross-check of results. Table~\ref{tab:agn} summarises flux measurements of the AGN. We can see that all three methods recover almost the same flux, and that the spectrum is relatively flat, as expected from an AGN. Furthermore, spectral index maps of ESO~137--006 obtained from uSARA and AIRI images (Figures~\ref{fig:sara}-\ref{fig:airi}, panel (g)) are in broad agreement with the findings of \citet{ramatsoku20} based on the CLEAN images {of ESO~137--006 at close frequencies}, where the lobes exhibit steep spectra, with a spectral index close to 5 at the tails.

\begin{deluxetable*}{ccccc}
\tablecaption{Flux measurements of the AGN in ESO~137--006 at both bands and its spectral index values, recovered by uSARA, AIRI, and CLEAN. \label{tab:agn}}
\tablecolumns{5}
\tablenum{1}
\tablehead{
\nocolhead{m}  &\colhead{}& \colhead{} &\colhead{} &\colhead{Frequency} \\
\nocolhead{m}  &\colhead{uSARA} &\colhead{AIRI} &\colhead{CLEAN}  &\colhead{(MHz)}
}
\startdata
Flux  &  $143$ & $143$ & $146$ &1053 \\
 (mJy) & $124$ & $125$ & $123$ &1399 \\
 \hline
 	Spectral index & $0.50$ & $0.47$ & $0.60$ &\\
\enddata
\tablecomments{{Flux measurements of CLEAN are computed from the convolved model images over a region centred at the AGN, and of the size of the main lobe of the associated dirty beams. The source being super-resolved in uSARA and AIRI images, its flux measurements are computed over its active pixels.}}
\end{deluxetable*}
The central region zooms (panels (b) and (e) of each figure) highlight both the super-resolution potential of uSARA and AIRI and the difficulty of interpreting features in RI images. At low band, both AIRI and uSARA recover what appear to be additional sets of filaments, or CSTs: $\textrm{T}_1$ and $\textrm{T}_2$, respectively North and South of the core/jets structure, $\textrm{T}_3$ further South , and $\textrm{T}_0$, a filament already detected by {\citet{ramatsoku20}}. Pixel intensity values of these filaments are within the range $[0.1,0.4]$~mJy, not only well above the image domain noise level of about 0.0014~mJy, but also at least 5 times higher than the level of the imaging artefacts, induced by the lack of amplitude calibration. The filaments are roughly parallel to the linear core/jets structure. This is a cause for both excitement and wariness. On the one hand, one explanation for the origin of CSTs is shearing of relativistic electrons off the jets, which then follow the ambient magnetic field with possibly stretched field lines \citep[see also][]{Condon2021}. From that point of view, additional inner filaments parallel to the jets make physical sense. On the other hand, the core/jets structure is one of the brightest features in the image, and one must always be wary of secondary image features that appear to trace bright features too closely, since calibration and deconvolution artefacts could easily take this form (being modulated by sidelobes of the dirty beam). One must therefore carefully compare reconstructions made with different methods and at different frequencies, since imaging artefacts tend to scale spectrally following the geometry of the dirty beam.

{The innermost filaments $\textrm{T}_1$ and $\textrm{T}_2$ appear in both the low and high band images made by AIRI and uSARA. $\textrm{T}_2$ is also confirmed by the CLEAN high band image and is not inconsistent with the CLEAN low band image, where there is emission blending with the core/jets. $\textrm{T}_1$ is not inconsistent with the CLEAN images either, where there is also emission, however not resolved at all. {This is also backed up by the associated model images (Figure~\ref{fig:wsclean}, panels (b') and (e')), where large scale components are recovered around the same region}. The first sidelobe of the dirty beam (indicated by dashed circles in panels (b) and (e) of each figure) 
is in any case only slightly smaller than the separation between the jets and $\textrm{T}_1$ and $\textrm{T}_2$, explaining why the filaments are not well resolved by CLEAN. Finally, the core/jets structure {recovered by AIRI and uSARA} has a clear discontinuity between the core and the jets, while the innermost filaments show no such thing. {For CLEAN, although the core and the jets are fully connected on the model images (Figure~\ref{fig:wsclean}, panels (b') and (e')), the connection seems weaker after convolution with the restoring beam (panels (b) and (e) of the same figure).} The position of the filaments shifts slightly (by about a pixel) between the low and high band images, but in the opposite direction than that which would be expected from a dirty beam-modulated artefact. This is possibly due to pixelation effects in the reconstructions. The recovered spectral indices are inconclusive. On balance, we must conclude that the innermost filaments are likely to be physical CSTs and not imaging artefacts. }

Although reconstructed by both uSARA and AIRI, the nature of $\textrm{T}_3$ is less conclusive, as it only appears at low band. It may be that, at high band, the reconstruction confuses it with the more complex structure of $\textrm{T}_0$, just to the South of it. We note that $\textrm{T}_0$ is reconstructed by all algorithms, with an impressively clear and finely resolved East-West connection at low band by uSARA and AIRI, while the structure is very blurred and interrupted in the CLEAN image.

Finally, examples of what are almost certainly artefacts are the fainter extended structures labelled $\textrm{S}$. They are roughly parallel to the radio galaxy structure, and scale inwardly with the dirty beam in the high band images. Since they are reconstructed by all three methods, they are likely to be residual amplitude calibration errors.

\paragraph{Computational cost.}
 Table~\ref{tab:compute-cost} summarises the computational cost of the imaging algorithms. Specific to uSARA and AIRI, the {computation} cost associated with the decomposition of the measurement operator is reported alongside the cost to run the imaging algorithm. As expected, with AIRI leveraging a fast denoiser on GPU and uSARA relying on a sub-iterative denoiser on CPU, the former brings a significant reduction of the imaging cost over the latter: about 2.3 times less CPU hours at both bands, with a negligible amount of GPU hours. AIRI was only 4 times more expensive than WSClean in the imaging process, and 7 when including the {computation} cost {of the measurement operator}. As AIRI denoisers are trained completely independently of the data to be imaged, the training cost associated with the single denoiser used for both bands is not considered part of the computational cost.

\begin{deluxetable*}{lchccc}
\tablecaption{Computational cost of uSARA, AIRI, and CLEAN in CPU hours (h). Specifically to AIRI, GPU h used by the DNN denoiser are also reported.\label{tab:compute-cost}}
\tablecolumns{6}
\tablenum{2}
\tablewidth{0pt}
\tablehead{
\colhead{} &\colhead{uSARA}& \nocolhead{}  &\colhead{AIRI} &\colhead{CLEAN} &\colhead{Frequency} \\
 \colhead{} & \colhead{(CPU h)} & \nocolhead{} &\colhead{(CPU h, GPU h)} & \colhead{(CPU h)} & \colhead{(MHz)}
}
\startdata
 {Computation {of $\Phib^\dagger \Phib$} }\tablenotemark{$*$}   & 427  &  &427, - & - &  1053 \\
  Imaging                           & 1120 &  &480, 5 &  132 & \\
 \hline
 {Computation of $\Phib^\dagger \Phib$} \tablenotemark{$*$} & 652  &  & 652, - &  -  &  1399   \\
 Imaging                           & 2377 &  & 1028, 6  & 236  & 
\enddata
\tablenotetext{*}{{Computation cost of the underlying sparse matrices}.}

\end{deluxetable*}

\section{Conclusions}
\label{sec:mainconcl}
We have introduced the first AI-based framework for deep, super-resolution, wide-field RI imaging, based on a plug-and-play scheme whereby a dedicated denoising operator is injected as an image regulariser in an optimisation algorithm. We have demonstrated two image reconstruction algorithms, uSARA and AIRI, respectively propelled by powerful handcrafted and learned denoisers, aiming at delivering a high level of imaging precision. Both algorithms are highly parallelised for scalability, via automated image faceting and decomposition of the RI measurement operator into sparse low-dimensional building blocks. {An in-depth study of practical scalability to the extreme data and image dimensions expected in the SKA context, in particular for wideband imaging, is warranted}. uSARA and AIRI were used to revisit MeerKAT L-band observations of a wide FoV containing ESO~137-006, from 19 gigabytes of visibility data. Our results confirm the ability of uSARA and AIRI to access a new regime of imaging resolution and dynamic range with respect to CLEAN. We have studied in particular the wealth of filamentary structure revealed within ESO~137--006’s radio lobes, some of which are likely CSTs. 
Our results also demonstrate further improvement brought by AIRI over uSARA in both dynamic range and speed, underpinned by the hybrid approach at the interface of optimisation and AI.

\section*{Data Availability}
The utilised data are observations with the MeerKAT telescope (Project ID SCI-20190418-SM-01). The images underlying this article are available from the research portal of Heriot-Watt University, with the digital object identifier \texttt{\href{https://researchportal.hw.ac.uk/en/datasets/unveiling-structure-in-eso-137006-with-airi-and-usara}{10.17861/b3a4ea6b-805c-4629-8c5c-fbb1f06ab53a}}.

\section*{Acknowledgements}
The research of AD, AJ and YW was supported by the UK Research and Innovation under the EPSRC grant EP/T028270/1 and the STFC grant ST/W000970/1. The work of OS was supported by the South African Research Chairs Initiative of the National Research Foundation (NRF), an agency of the Department of Science and Innovation (DST). The research of MR is supported by the New Scientific Frontiers grant of the South African Radio Astronomy Observatory (SARAO). The research used Cirrus, a UK National Tier-2 HPC Service at EPCC funded by the University of Edinburgh and EPSRC (EP/P020267/1). The MeerKAT telescope is operated by SARAO, which is a facility of DST/NRF.

\bibliography{AI4RI}{}
\bibliographystyle{aasjournal}

\appendix
\section{The Forward-Backward algorithmic structure}\label{sec:fb}
In this work, the inverse RI imaging problem is approached as an optimisation problem. In the context of optimisation theory, an ``objective function'' is defined, typically as the sum of a data fidelity term $f$ and a regularisation term $r$ injecting a prior image model to compensate for data incompleteness. The image estimate is defined as the minimiser of this objective and is reached via provably convergent algorithms \citep{bauschke2017convex}. The obtained solution can also be understood in a Bayesian framework as a \emph{maximum a posteriori} (MAP) estimate with respect to a posterior distribution, the negative logarithm of which is the objective. More specifically to our setting, we aim at solving
\begin{equation}
\label{eq:objfun}
 \underset{\xb\in \eR^N}{\mathrm{minimise}} ~~ f(\xb; \yb)+ \lambda r(\xb).
\end{equation}
In this objective, $f$ is a convex Lipschitz-differentiable function of a variable $\xb\in \eR^N$ representing the image variable, also a function of the data vector $\yb \in \eC^M$, whose role is to enforce fidelity to data. The function $r$ is a convex and possibly non-differentiable function of $\xb$, encoding the prior image model. The regularisation parameter $\lambda >0$ acts as a trade-off between the two terms. Problems of the form (\ref{eq:objfun}) can be solved via the iterative FB algorithm, alternating between a forward step in the negative direction of the gradient of $f$, and a backward step involving a simple denoising operator, known as the proximal operator of the regularisation function $r$. The proximal operator is itself defined as the solution of a (simpler) minimisation problem: 
$\operatorname{prox}_{\lambda r}(\zb) = \text{argmin}_{\ubb\in \eR^N} \lambda r(\ubb)+\|\zb-\ubb\|_2^2/2$, for any $\zb \in \eR^N$ and $\lambda>0$. The proximal operator of simple functions $r$ often benefits from a closed-form solution (\emph{e.g.} the proximal operator of the $\ell_1$ norm is a simple component-wise soft-thresholding operator). However, proximal denoisers of sophisticated regularisations must usually be computed iteratively, as solutions of the minimisation task by which they are defined.

The FB iterative structure reads
\begin{equation}
\label{eq:fb}
 (\forall k\in \eN) \qquad \xb^{k+1} = \operatorname{prox}_{\gamma \lambda r}( \xb^{k} -\gamma \nabla f (\xb^{k} )),
\end{equation}
where the step-size $\gamma>0$ is strictly upper-bounded by $2/L$ to ensure convergence, with $L$ being the Lipschitz constant of $\nabla f$. 

Interestingly, the recently emerged PnP scheme has established that proximal optimisation algorithms such as FB, not only enable the use of proximal operators of handcrafted regularisation functions, but also the injection of learned DNN denoisers defining the regularisation term implicitly \citep{venkatakrishnan2013plug, romano2017little}. We note that, in order to preserve algorithm convergence, and interpretability of its solution, the PnP denoiser must typically satisfy a ``firm non-expansiveness'' constraint, ensuring that it contracts distances \citep{Pesquet21, hurault2022proximal}.

Our RI imaging framework relies on a data fidelity term which reflects the Gaussian nature of the noise, and is given by
$f(\xb; \yb)=1/2\|\Phib\xb-\yb\|_2^2$, where $\|\cdot\|_2$ denotes the standard $\ell_2$ norm. 
Its gradient reads $\nabla f (\xb) = \text{Re}\{ \Phib^\dagger\Phib\} \xb -\text{Re}\{\Phib^\dagger\yb \}$, where $(\cdot)^\dagger$ denotes the adjoint of its argument. The Lipschitz constant of $\nabla f $ is given by $L=\|\text{Re}\{\Phib^\dagger\Phib\}\|_{\textrm{S}}$, where $\|\cdot\|_{\textrm{S}}$ denotes the spectral norm. 
As for the image regularisation, we leverage the versatility of the PnP framework and investigate both advanced handcrafted and learned prior image models, respectively propelling the uSARA and AIRI imaging algorithms. 
\vspace*{-0.1cm}

\section{uSARA's handcrafted denoiser}
\label{ssec:sara}
uSARA's image model, originally proposed in \citet{Carrillo2012}, promotes the non-negativity of the intensity image and its sparsity in an overcomplete dictionary $\Psib\in\eR^{N\times B}$, which consists of a normalised concatenation of orthogonal wavelet bases. The sparsity model is encoded via a non-differentiable log-sum regularisation function $r$, generalising the $\ell_1$ norm. The resulting multi-term regularisation thus reads \citep{repetti21, thouvenin2022parallel1}
\begin{equation}
 \lambda r(\xb) = \lambda \sum_{n=1}^{B} \rho \log ( 1 + | (\Psib^\dagger \xb )_n |/\rho ) + \iota_{\eR^N_+}(\xb),
\end{equation}
where $(.)_n$ denotes the $n^{\textrm{th}}$ coefficient of its argument vector, and $\iota_{\eR_+^N}$ denotes the indicator function of the real positive orthant, imposing the non-negativity constraint: $\iota_{{\eR^N_{+}}} (\xb)=+\infty$ if $\xb \notin {\eR^N_{+}}$ and 0 otherwise. The log-sum regularisation being non-convex, the minimisation task is approached via a re-weighting procedure where a series of convex surrogate minimisation tasks, composed of weighted-$\ell_1$ regularisation with non-negativity constraint, are solved using FB \citep{repetti20, repetti21, terris22}. The denoising proximal operator of the resulting multi-term regularisation does not have a closed-form solution and is solved iteratively. In other words, uSARA relies on FB with sub-iterative regularisation denoisers. 

We note that the parameter $\lambda$ controls a soft-thresholding operation acting on the wavelet coefficients. As proposed by \citet{terris22}, the exact thresholding parameter $\gamma\lambda$ is set equal to the measurement noise transferred to the image domain $\tau/\sqrt{2 L}$\footnote{When considering a data weighting scheme other than natural weighting, such as uniform or Briggs weighting, a multiplicative correction factor is applied to $L$ for a more accurate noise estimate in the image domain \citep{wilber221}.}:
\begin{equation}
\label{eq:lambda}
 \gamma\lambda={\tau}/{\sqrt{2 L}},
\end{equation}
with the step-size $\gamma$ typically set to $\gamma=1.98/L$.
The parameter $\rho>0$ represents a floor level on the wavelet coefficients and is set naturally to the noise level $\rho=\gamma\lambda$ \citep{thouvenin2022parallel1}.

\section{AIRI's learned denoiser}
\label{ssec:airi}
Following \cite{terris22}, we trained a convolutional DNN, with a simple DnCNN architecture \citep{zhang2017beyond}, on a rich, synthetic database $\mathcal{U}$ of normalised images with adaptive dynamic range. The training loss is a classical $\ell_1$ loss, enhanced with a firm non-expansiveness constraint on the denoiser:
\begin{equation}
\begin{aligned}
 &\underset{\operatorname{D}}{\text{minimise}}\,\, \mathbb{E}_{{\ubb}\sim \mathcal{U}, {\nb}\sim\mathcal{N}(0,1)}\Big[ \|\operatorname{D}({\ubb}+\sigma {\nb})-{\ubb} \|_1\Big] \\
 &\text{such that}\,\,(\forall {\ubb}\in\mathbb{R}^N) \,\, \|\nabla_{{\ubb}}\!\left(2\operatorname{D}-\mathbb{I}\right)\|_{\textrm{S}} \leq 1,
\end{aligned}
\label{eq:training_loss}
 \end{equation}
where $\ubb\in\mathbb{R}^N$ are samples of the training database $\mathcal{U}$, $\nb \sim \mathcal{N}(0,1)$ is an additive Gaussian random noise, $\sigma>0$ is the training noise level, $\mathbb{E}$ denotes the expectation taken over $\ubb$ and $\nb$, $\mathbb{I}$ denotes the identity operator, and $\|\cdot\|_1$ denotes the $\ell_1$ norm. 

We emphasise that training under the firm non-expansiveness constraint is a highly challenging task. As proposed in \citet{Pesquet21}, in practice, we relax the constraint and introduce a variant of the regularisation in (\ref{eq:training_loss}), which penalises softly non firmly nonexpansive networks. We further note that, while \citet{terris22} demonstrated in simulation that this leads to a robust way to ensure convergence of the resulting PnP algorithms, we have witnessed that, when used for real data and at large image sizes and dynamic ranges such as those of interest here, some denoisers lead to algorithm instability, requiring further training.

The performance of the learned image regularisation is highly dependent on the training noise level $\sigma$, the impact of which mirrors that of the regularisation parameter $\lambda$ in uSARA's denoiser. \citet{terris22} proposed a heuristic according to which $\sigma$ should be set equal to the measurement noise transferred to the image domain $\tau/\sqrt{2 L}$. However, the training database is normalised, with peak image values upper-bounded by $1$. To avoid any generalisability issues, the trained denoisers should therefore be used on similarly normalised images, which was the case for the test images in the simulation framework of \citet{terris22}. In general, this constraint can be accommodated by rescaling the inverse problem (\ref{eq:invpb}), effectively dividing it by an upper bound on the peak intensity of the sought image, $\alpha\geq \operatorname{max}_j\{{x}_j\}$, which can be inferred from the peak of the dirty image. The rescaled inverse problem,
\begin{equation}
{\yb}/{\alpha} = \bm{\Phi}\left({\overline{\xb}}/{\alpha}\right)+{\overline{\nb}}/{\alpha},
\label{eq:invpb2}
\end{equation}
now targets the recovery of ${\overline{\xb}}/{\alpha}$, with a peak value upper-bounded by 1. As a result, the heuristic generalises to setting $\sigma$ equal to the inverse input image-domain peak signal-to-noise ratio, which can be understood as the target reconstruction dynamic range, rather than an absolute noise level:
 \begin{equation}
\sigma={\tau}/{\alpha \sqrt{2L}}.
\label{eq:sigma}
\end{equation}
Interestingly, if a pre-defined denoiser, trained at some high dynamic range, is available, any RI dataset with signal-to-noise ratio \emph{a priori} pointing to a lower dynamic range denoiser, can be further rescaled (with a larger $\alpha$) to match the existing denoiser, according to (\ref{eq:sigma}). We adopt here this single denoiser approach, where different data are matched to the denoiser rather than the contrary. Naturally, PnP solutions are multiplied by $\alpha$ after reconstruction.

\section{Denoiser faceting}
\label{app:denoiser}
Specific to the algorithm scalability requirement, arising from the large image dimensions of interest, we propose an automated parallelisation of the studied denoisers, enabled by image faceting. Firstly, uSARA's proximal denoiser takes advantage of a faceted implementation of the sparsity dictionary $\Psib$, enabled by its convolutional nature and the compact support of the wavelet kernels \citep{Prusa2012}. The number of facets is set to optimise the parallelisation of the processing across the available CPUs under communication constraints. 
Secondly, AIRI's learned denoiser is decomposed and applied independently across facets of the image. This procedure is enabled by the convolutional nature of the DNNs, which rely on kernels with compact support, in turn yielding a small receptive field \citep{luo2016understanding}. The number of facets is set to optimise the parallel processing across the available GPUs for scalability, under memory constraints.

\section{Parallel wide-field measurement operator}
\label{sec:mo}
Firstly, on large FoV such as the one of interest here, the $w$ component of the baselines induces a non-negligible baseline-dependent chirp-like phase modulation on the radio sky \citep{Cornwell2005, Wiaux09}. This $w$-effect can be formulated in closed form and
needs to be accounted for in the model of the measurement operator. Its modelling as a simple phase modulation in the image domain for each baseline is however impractical when used in combination with the Fast Fourier Transform (FFT) underpinning the fast implementation of $\Fb\in \eC^{N^\prime \times N^\prime}$ in (\ref{eq:invpb}), which computes all the discrete coefficients of the Fourier plane at once rather than a selected $(u,v)$ point.  For accurate and computationally efficient modelling, we consider a hybrid approach combining the $w$-stacking \citep{Offringa2014} and the $w$-projection approaches \citep{Cornwell2005}, whereby the measurements are grouped into $P$ $w$-stacks composed of $M_p$ data points each, with $1\leq p \leq P$, resulting from binning the visibilities in the $w$ dimension. The $w$-modulation of each visibility is decomposed into two components: (i) a large phase modulation associated with the central $w$ value of the $w$-stack to which it belongs, incorporated in the measurement operator through phase modulation in the image domain, and (ii) an offset phase modulation injected through convolution with a small $w$-kernel \citep{dabbech17} in the Fourier plane. The resulting measurement operator $\Phib$ is decomposed into a series of sparse operators as
\begin{align}
\label{eq:phip}
    \Phib &= \begin{bmatrix}
           {\Phib}_1 \\
           {\Phib}_2 \\
           \vdots \\
           {\Phib}_P
         \end{bmatrix}
  \end{align}
where the operator ${\Phib}_p= \Gb_p \Fb \Zb_p\in \eC^{M_p \times N}$ is the measurement operator associated with the $p^{\textrm{th}}$ $w$-stack. More specifically, $\Zb_p \in \eC^{N^\prime \times N}$ denotes the zero-padding operator which encompasses the $w$-modulation of the associated $w$-stack, in addition to the correction for the convolution with the approximate non-uniform Fourier transform interpolation kernels, and $\Gb_p\in \eC^{M \times N^\prime}$ is the sparse de-gridding matrix, encoding row-based convolutions between these kernels and the small $w$-kernels implementing the phase modulation of the $w$-offsets in the Fourier plane. Combining $w$-stacking and $w$-projection results in a memory-efficient, and accurate measurement operator. We also emphasise that any DDE calibration solutions modelled as Fourier kernels can be easily injected in the measurement operator model via further row-based convolution \citep{dabbech21}.

Secondly, the operator of interest in the gradient step of the FB iterative structure (\ref{eq:fb}) is not ${\Phib\in \eC^{M \times N}}$ but rather ${\Phib^\dagger \Phib \in \eC^{N \times N}}$, which now reads
\begin{equation}
\label{eq:phidagphi}
 {\Phib^\dagger \Phib}=\sum_{p=1}^P {\Phib}^\dagger_p {\Phib}_p, 
\end{equation}
with ${\Phib}^\dagger_p {\Phib}_p =\Zb_p^\dagger\Fb^\dagger \Hb_p \Fb \Zb_p \in \eC^{N \times N}$, and where the holographic matrices $\Hb_p= \Gb^\dagger_p \Gb_p\in \eC^{N^\prime \times N^\prime}$ encode both the de-gridding and gridding steps. {The scalability of ${\Phib^\dagger \Phib}$ to large data acquisition regimes is promoted by enabling three key features.}

{Firstly, a dimensionality reduction feature to reduce the memory requirements of ${\Phib}^\dagger {\Phib}$ is supported. The functionality consists in gridding the data and encoding the de-gridding and gridding steps as a single operation with the holographic matrices $\Hb_p$, directly implemented as sparse operators. By doing so, the operator ${\Phib^\dagger \Phib}$ becomes effectively blind to the data dimension $M$.}

{Secondly, a planning strategy to automate the choice of the number of the $w$-stacks and the decision to enable the dimensionality reduction from a subset of the data is devised. In the first instance, estimates of the computational complexity of the application of ${\Phib}^\dagger {\Phib}$ (derived from the number of FFTs and the sparsity of the de-gridding matrices) and of the memory required to host the de-gridding matrices are obtained for a wide range of values of the $w$-stacks number. The retained value is the one presenting the best trade-off between the computational cost and the memory requirements,
under constraints set by the computing architecture on which the imaging algorithm is deployed (number of compute nodes, number of CPUs per node, available memory per CPU, etc). Data dimensionality reduction via visibility gridding is enabled when the memory requirements exceed the available resources.}

{Thirdly, a fully automated parallelisation of ${\Phib^\dagger \Phib}$ is achieved through memory-based partitioning of its underlying de-gridding/holographic matrices and data vectors. The sparse matrices are computed as part of the initialisation of the imaging algorithm. A data-clustering step is first performed in parallel for each $w$-stack to further distribute its de-gridding/holographic matrix into blocks. The clusters are made of visibilities belonging to the same radial slice of the Fourier plane, minimising the amount of underpinning discrete Fourier coefficients and subsequent communication requirements. The angular opening of each radial slice is determined by the memory needed to compute and host the resulting blocks. 
From the identified number of clusters, the CPUs dedicated to the forward step are allocated. The blocks of the de-gridding/holographic matrices are then computed only once and hosted directly on the compute nodes, to be applied in parallel at each FB iteration.}

\section{Automated parameter selection}
\label{app:params}
The estimated image noise levels at the low and high bands are respectively $\tau/\sqrt{2 L}\simeq 0.0014$~mJy and $\tau/\sqrt{2 L}\simeq 0.0017$~mJy. The peak intensity values, as estimated from the normalised\footnote{The dirty images are normalised by $\beta=\max_i(\text{Re}\{\Phib^\dagger{\Phib}\}\boldsymbol{\delta})_{i} $, where $\boldsymbol{\delta}$ is an image with value 1 at the phase centre and 0 otherwise. By doing so, the dirty beam $\text{Re}\{\Phib^\dagger{\Phib}\}{{\boldsymbol{\delta}}}/\beta$ has a peak value equal to 1.} dirty images, are 0.69~Jy at low band and 0.37~Jy at high band. For uSARA, $\gamma \lambda$ and $\rho$ are set equal to the estimated noise levels, as per (\ref{eq:lambda}). For AIRI, the estimated noise and peak values suggest target dynamic ranges of $5\times 10^5$ and $2.2\times 10^5$ at low and high bands, respectively. Owing to the chosen normalisation of the dirty image for peak estimation, the dirty peak value consistently overestimates the true peak value, so that the real target dynamic ranges are below these values\footnote{In fact, the peak values reconstructed by uSARA and AIRI are around 0.05~Jy for both bands, suggesting an initial overestimation with the normalised dirty image by more than a factor 10. Note that this uncertainty in the initial estimate is not a problem as only an upper bound is required.}. In this context, we have used a single denoiser trained for target dynamic range $4 \times 10^5$, rescaling the inverse problems by the appropriate $\alpha$ at each band independently as in (\ref{eq:invpb2}). In other words, after rescaling, and as per (\ref{eq:sigma}), the inverse problem at each band is affected by a noise of standard deviation $\tau/\alpha \sqrt{2 L}$ equal to the training noise level of the chosen denoiser, \emph{i.e.}~$\sigma=2.5 \times 10^{-6}$~Jy.

uSARA and AIRI denoisers were respectively applied on $8\times8$ and $2\times 2$ facet decompositions of the images. The low and high bands data were decomposed into 12 and 14 $w$-stacks in (\ref{eq:phip}), with $\Phib^\dagger \Phib$ encoded via the holographic matrices $\Hb_p$. At low and high bands respectively, this enabled to lower memory requirements from 470 and 645 gigabytes needed to host the de-gridding matrices $\Gb_p$, down to 81 and 159 gigabytes. Finally, in WSClean, multiscale CLEAN utilised 72 $w$-stacks for both bands. 



\end{document}